\documentclass[twocolumn,aps,prl,superscriptaddress]{revtex4-1}
\pdfoutput=1
\usepackage[pdftex]{graphicx}
\usepackage{color}
\usepackage{amsmath}
\usepackage{braket}
\usepackage[latin9]{inputenc}
\usepackage[T1]{fontenc}
\usepackage{url}
\usepackage[english]{babel}
\usepackage{graphicx}% Include figure files
\usepackage{caption}
\usepackage[export]{adjustbox}% to align figures
\usepackage{dcolumn}% Align table columns on decimal point
\usepackage{bm}% bold math
\usepackage[caption=false]{subfig}
\usepackage{float}
\usepackage{url}
\usepackage{verbatim}

\newcommand{\be}{\begin{equation}}
\newcommand{\ee}{\end{equation}}
\newcommand{\bea}{\begin{eqnarray}}
\newcommand{\eea}{\end{eqnarray}}
\newcommand{\non}{\nonumber}
\newcommand{\bi}{\begin{itemize}}

\newcommand{\ei}{\end{itemize}}
\newcommand{\bn}{\begin{enumerate}}
\newcommand{\en}{\end{enumerate}}

\def\s#1{{\scriptscriptstyle #1}}

\def\1eq#1{Eq.~(\ref{#1})}
\def\2eqs#1#2{Eqs.~(\ref{#1}) and~(\ref{#2})}
\def\3eqs#1#2#3{Eqs.~(\ref{#1}), (\ref{#2}) and~(\ref{#3})}
\def\4eqs#1#2#3#4{Eqs.~(\ref{#1}), (\ref{#2}), (\ref{#3}) and~(\ref{#4})}
\def\noeq#1{(\ref{#1})}

\def\fig#1{Fig.~\ref{#1}} 

\begin{document}

\title{The lattice gluon propagator in renormalizable $\xi$ gauges} 

\author{P.~Bicudo} 
\affiliation{CFTP, Instituto Superior T\'{e}cnico, Universidade de Lisboa, Av. Rovisco Pais, 1049-001 Lisboa, Portugal}

\author{D.~Binosi}
\affiliation{European Center for Theoretical Studies in Nuclear Physics and Related Areas (ECT*) and Fondazione Bruno Kessler, Italy}

\author{N.~Cardoso} 
\affiliation{NCSA, University of Illinois, Urbana IL 61801, USA}

\author{O.~Oliveira} 
\affiliation{CFisUC, Department of Physics, University of Coimbra, P-3004 516 Coimbra, Portugal}

\author{P.~J.~Silva}
\affiliation{CFisUC, Department of Physics, University of Coimbra, P-3004 516 Coimbra, Portugal}

\begin{abstract}
We study the SU(3) gluon propagator in renormalizable $R_\xi$ gauges implemented on a symmetric lattice with a total volume of (3.25 fm)$^4$ for values of the guage fixing parameter up to $\xi=0.5$. As expected, the longitudinal gluon dressing function stays constant at its tree-level value $\xi$. Similar to the  Landau gauge, the transverse $R_\xi$ gauge gluon propagator saturates at a non-vanishing value in the deep infrared for all values of $\xi$ studied. We compare with very recent continuum studies and perform a simple analysis of the found saturation with a dynamically generated effective gluon~mass.
 \end{abstract}

\maketitle

{\it Introduction.} Dynamical mass generation is arguably one of the most important non perturbative features of QCD. Within the quark sector, this phenomenon originates from a cloud of low-momentum gluons attaching themselves to the current quark.  At the theoretical level this effect is described by a so-called (quark) gap equation; its solutions describe the evolution of a (chiral) current quark of perturbative QCD into a constituent quark (with a mass $\sim$ 350 MeV) as the momentum flows from UV to IR values. This dynamically generated mass accounts alone for the largest part of the proton's mass, explaining how hadron masses emerge (dynamically) in a universe with light quarks. 

In the gauge sector, the propagation of the aforementioned gluons is also described by a gap equation. In the Landau gauge, its solutions saturate in the IR to a non-vanishing value. This can be interpreted, as in the quark case, through the concept of a dynamically generated (and momentum dependent) mass~\cite{Cornwall:1981zr,Bernard:1982my,Donoghue:1983fy,Philipsen:2001ip,Aguilar:2004sw,Oliveira:2010xc,Aguilar:2011ux,Binosi:2012sj,Aguilar:2014tka}. The associated gluon mass function $m^2$ is a monotonically decreasing function that appears in the transverse part of the propagator (hence gauge invariance is not tampered with); in addition, it is power-law suppressed in the UV (and therefore unobservable in perturbative applications). 

The dynamical generation of a gluon mass substantiate the hope that QCD is nonperturbatively well defined, implying that the theory generates by itself the needed IR cutoffs. In addition, its presence affects the theory in a variety of ways: for example, it gives rise to a quark-gluon interaction coinciding with that required to describe ground-state observables~\cite{Binosi:2014aea}.   

Lattice gauge theories have been instrumental in unequivocally establishing IR saturation in the gauge sector~\cite{Cucchieri:2007md,Cucchieri:2010xr,Bogolubsky:2007ud,Bogolubsky:2009dc,Oliveira:2009eh,Ayala:2012pb,Oliveira:2010xc,Oliveira:2012eh}. In fact, once discretization artefacts are accounted for, their results are equivalent to exact all-order results, and therefore are regarded as benchmark tests for the different continuum approaches~~\cite{Boucaud:2008ky,Aguilar:2008xm,Fischer:2008uz,RodriguezQuintero:2010wy,Pennington:2011xs,Campagnari:2010wc,Alkofer:2000wg,Maris:2003vk,Aguilar:2004sw,Fischer:2006ub,Kondo:2006ih,Braun:2007bx,Binosi:2007pi,Binosi:2008qk,Kondo:2011ab,Szczepaniak:2001rg,Szczepaniak:2003ve,Epple:2007ut,Szczepaniak:2010fe,Watson:2010cn,Watson:2011kv,Rojas:2013tza}. While, in general, lattice computations do not need any gauge fixing (GF), the latter becomes essential for the computation of Green's functions (propagators and vertices), as it is well-known that these objects depend on the gauge condition employed. This is not a problem {\it per se}, as there are many instances in which physics is better understood in a particular gauge ({\it e.g.}, partons make sense only in the light cone gauge); however, it clearly highlights the importance of performing simulations in as many gauges as possible, in order to discern which aspects of the nonperturbative behavior of the function at hand are (or are not) affected by a gauge choice. 

While the gluon two-point function has been studied in covariant and non-covariant gauges~\cite{Cucchieri:2006hi,Bornyakov:2003ee,Mendes:2006kc,Cucchieri:2007uj}, reliable calculations in renormalizable-$\xi$ ($R_\xi$) gauges~\cite{Fujikawa:1972fe} have not been systematically pursued so far. Despite being the only class of gauges which are completely under control at the perturbative level, its lattice implementation has in fact proven to be quite complicated
\cite{Giusti:1996kf,Giusti:1999wz,Giusti:1999im,Giusti:1999cw,Giusti:2000yc,Giusti:2001kr}, 
due to a {\it no go} result~\cite{Giusti:1996kf}. A viable lattice formulation was finally put forward in Ref. \cite{Cucchieri:2009kk}. In practice, however, one still encounters significant convergence problems \cite{Cucchieri:2010ku,Cucchieri:2011aa,Cucchieri:2011pp} when attempting numerical GF, which unfortunately become more severe as the GF parameter $\xi$ and/or the lattice volume become larger, and the number of colors $N_c$ and/or the lattice coupling $\beta$ are small.  As a result, there have been only preliminary studies of the $R_\xi$ gluon propagator~\cite{Cucchieri:2009kk,Cucchieri:2011pp}.

In this letter, we present the SU(3) gluon propagator in $R_\xi$ gauges for a relatively large lattice volume (3.25 fm)$^4$ and a GF parameter up to $\xi=0.5$. This allows us to address in some detail the IR behavior of the $R_\xi$ gluon propagator, and study the dynamical generated gluon mass beyond the Landau gauge limit.

{\it $R_\xi$ gauges framework.} In the continuum, gauge fixing is achieved by adding to the SU($N_c$) Yang-Mills action the term (in Minkowski space),
\begin{equation}
 S_\s{\mathrm{GF}}=\int\!\mathrm{d}^4x\,\left[b^m \Lambda^m-\frac\xi2(b^m)^2\right].
\end{equation}
Here $\xi$ is a (non-negative) GF parameter, $b^m$ are the so-called Nakanishi-Lautrup multipliers and  $\Lambda^m=\Lambda^m[A]$ is the GF condition. For all fields we write $\Phi=\Phi^m t^m$, where $t^m$ are the SU($N_c$) generators. Going on-shell with the $b$ fields, one obtains the condition $\xi b^m=\Lambda^m$ and the corresponding GF action,
\begin{equation}
	S_\s{\mathrm{GF}}={ 1 \over 2 \xi} \int\!\mathrm{d}^4x\,( \Lambda^m)^2.
	\label{GFaction}
\end{equation}
$R_\xi$ gauges are obtained when the linear condition $\Lambda^m=\partial^\mu A^m_\mu$ is chosen. In this case the (non-perturbative) gluon propagator can be decomposed according to
\begin{equation}  
	\Delta_{\mu\nu}(q)=(g_{\mu\nu}-q_\mu q_\nu/q^2)\Delta_\s{\mathrm{T}}(q^2)+(q_\mu q_\nu/q^2)\Delta_\s{\mathrm{L}}(q^2)\,.
  \label{Rxiprop}
\end{equation}
Slavnov-Taylor identities ensure that $q^2\Delta_\s{\mathrm{L}}=\xi$ to all orders and, therefore, all the dynamical information is carried by the transverse form factor $\Delta_\s{\mathrm{T}}$.

The lattice formulation of Yang-Mills theories is obtained in terms of the Wilson gauge action, in which the dynamical variables are the gauge links $U_\mu$, related to the gauge fields (in lattice units) through,
\begin{align}
	U_{\mu}(x)&= \exp[ i g_0 A_\mu (x +  \widehat e_\mu / 2)]\, ,
 \non \\
  A_\mu (x +  \widehat{e}_\mu / 2 ) &= \left. \frac{U_{\mu}(x) - U^\dagger_{\mu}(x)}{2i g_0} \right|_{\mathrm{traceless}},
\label{link}
\end{align}
where 
%for simplicity we do not include $g_0$ in the normalization of $A_\mu$, 
$\widehat{e}_\mu$ is the unit vector along the direction $\mu$, $\beta=2N_c/g_0^{2}$ is the lattice coupling which determines the lattice spacing $a$. Physical quantities are then obtained by the evaluation of the Euclidean path integral through Monte Carlo techniques, with a probability distribution given by the exponential of the action.

In $R_\xi$ gauges, besides the usual integration over the link variables, one has to integrate over the $\Lambda$ fields. \1eq{GFaction} implies that the integration measure is a Gaussian distribution,
with variance $ \xi  $,
\begin{align}
	P\left[\Lambda^m(x)\right] \propto \exp \left\{- { 1 \over  2 \xi } \sum_m\,\left[\Lambda^m(x) \right]^2 \right\}.
	\label{gaussian}
\end{align} 
The numerical difficulty of implementing the $R_\xi$ gauges lies in enforcing the GF condition.

In fact, the standard procedure for GF requires to gauge rotate all link variables through the gauge transformation $ U_{\mu}(x) \rightarrow g(x)U_{\mu}(x)g^{\dag}(x+\widehat e_{\mu})$, where $g$ are elements of the SU($N_c$)  gauge group that minimizes a suitable functional implementing the desired GF condition. In the Landau gauge case, which is the $\xi\to0$ limit of the $R_\xi$ gauges studied here, the functional is,
\begin{align}
	{\cal E}_\s{\mathrm{LG}}[U,g] = -\mbox{Re} \, \mbox{Tr}\, \sum_{x, \mu}g(x) U_{\mu}(x) g^{\dagger}(x+ \widehat e_\mu)\,,
 	\label{gfLandau} 
\end{align}
which directly leads to the condition \mbox{$\nabla\!\cdot\! A^m=0$}.
Contrary to this simple limit, the general case of a non-vanishing~$\xi$ was proven to have no suitable GF functional to minimize~\cite{Giusti:1996kf}. This {\it no-go} theorem has been evaded in~\cite{Cucchieri:2009kk}, where it was shown that the functional,
\be
 {\cal E}_\s{R_\xi}[U,g]={\cal E}_\s{\mathrm{LG}}[U,g] +\mbox{Re} \, \mbox{Tr}\, \sum_x i g(x)\Lambda(x)\,,
 \label{Rxifunct}
\ee
yields the correct condition~$\nabla\!\cdot\! A^m=\Lambda^m$. In practice, the gauge transformation  $g$ is built as a product of a sequence of infinitesimal gauge transformations $g = \prod_j \delta g_j$. For each infinitesimal transformation $\delta g_j$ one minimizes the functional~\noeq{Rxifunct}; however when moving on to the next infinitesimal transformation $\delta g_{j+1}$,  the Gaussian distribution $\Lambda^m$ is maintained unchanged and  the link $U_\mu$ is updated through a gauge rotation. Writing $\delta g_j = 1+ i \sum_m w^m \,  t^m$, the variation of the functional~\noeq{Rxifunct} with respect to the coefficients $w^m$ reads,
\begin{align}
 & {\cal E}_\s{R_\xi}[U,\delta g]- {\cal E}_\s{R_\xi}[U,1] = \text{Tr} \sum_{x , \, m}  w^m(x) \, t^m\, \Delta (x) \ , 
	\\ \non 
 & \Delta(x) = \sum_\mu g_0 \Big[ A_\mu (x + \hat{e}_\mu / 2 )- A_\mu (x -  \hat{e}_\mu / 2 ) \Big] - \Lambda(x) \ .
\label{omega}
\end{align}
Thus, choosing~$w^m = \alpha \Delta^m$, with $\alpha$ is a relaxation parameter to be optimized, will reduce $\Delta$. Our goal is to converge to a vanishing $\Delta$ in all lattice points $x$, which implies,
\be 
  \theta = \frac{1}{ N_c L^4}\sum_x \mathrm{Tr}\, [\Delta(x) \Delta^{\dagger}(x)]\to0 \ .
\label{theta}
\ee
Whenever this condition is fulfilled (which, based on the experience of Landau GF  ~\cite{Cardoso:2012pv}, means to have $\theta < 10^{-15}$ ), then the configuration is $R_\xi$ gauge fixed. 

%---------------------------------------------------------------------------------------------------------------------
\begin{figure}[!t]
	\includegraphics[scale=0.65]{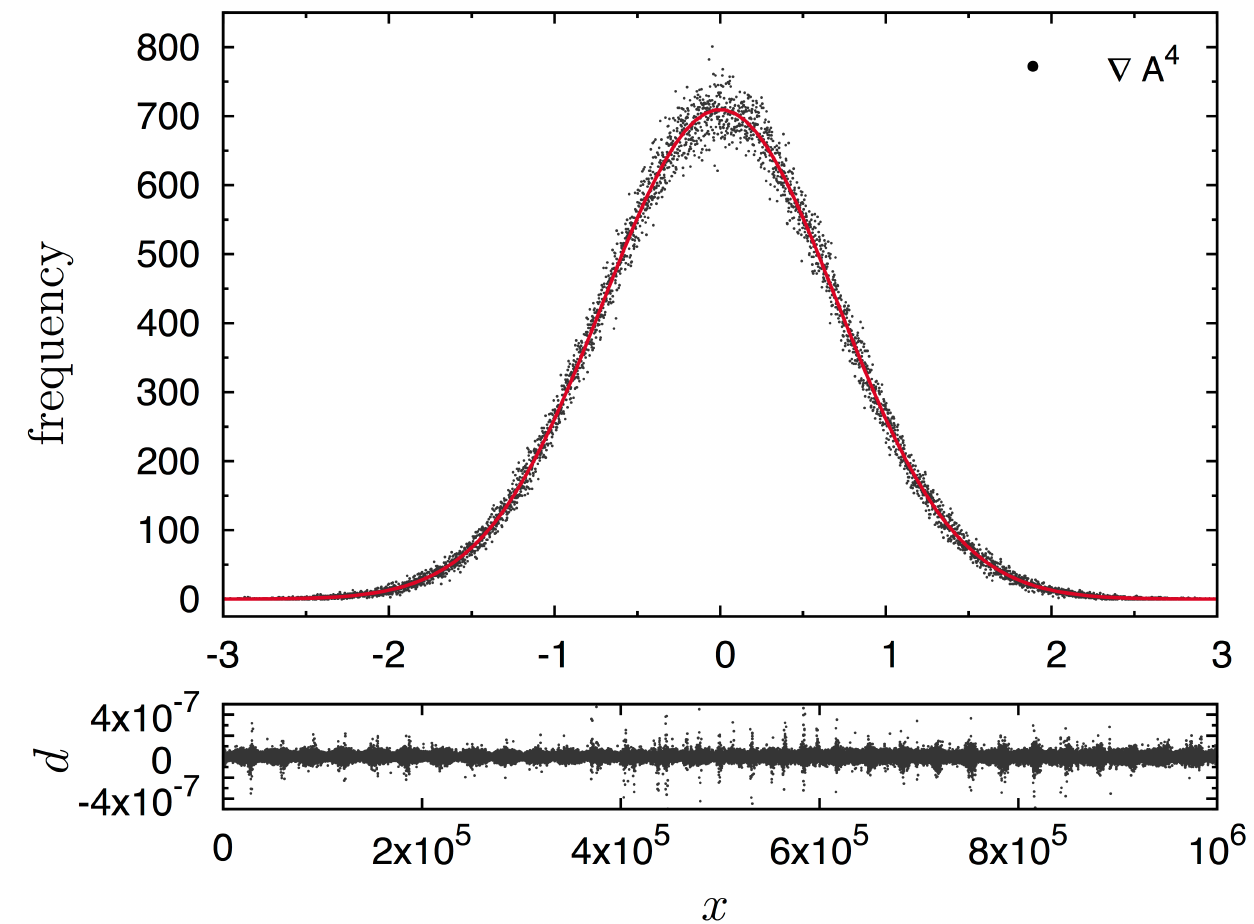}
	\caption{\label{fig:histo} 
	({\it Top}) The $32^4$ values of $\nabla\! \cdot\! A^4$ evaluated for a configuration gauge fixed at $\xi=0.5$, grouped in 5000 bins,  compared with a Gaussian  with standard deviation $\sqrt{\xi}\simeq0.316$. ({\it Bottom}) Plot of $d=\nabla\! \cdot\! A^4-\Lambda^4$; the two distributions coincide within $\sqrt\theta$ precision.}
\end{figure}

%---------------------------------------------------------------------------------------------------------------------
\begin{figure*}
\includegraphics[scale=1]{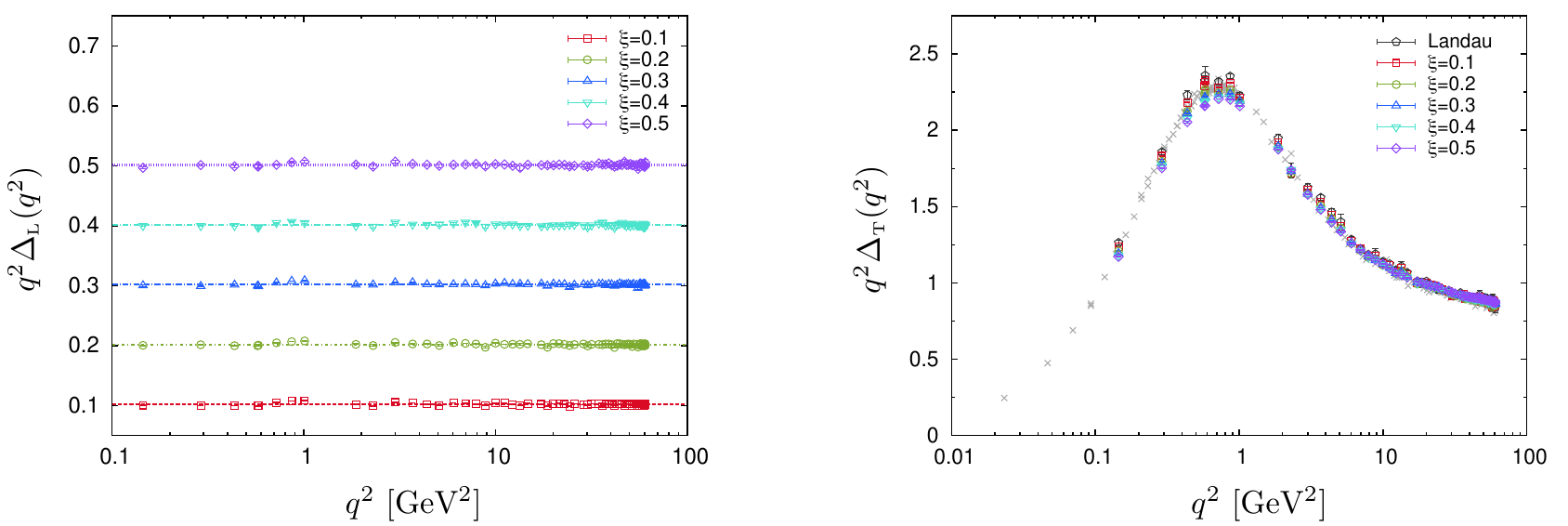}
\caption{\label{fig:gl-longit-trans}
({\it Left}) 
The $R_\xi$ longitudinal dressing function $q^2\Delta_\s{\mathrm{L}}\equiv\xi$; a fit of the data to a constant yield $\xi=$ 0.103(2), 0.203(2), 0.302(3), 0.402(3) and 0.502(3) respectively. 
({\it Right}) 
The $R_\xi$ gluon transverse dressing function $q^2\Delta_\s{\mathrm{T}}$. Landau gauge results obtained for a symmetric lattice of $L=80$ and $\beta=6.0$ (gray crosses) are also plotted~\cite{Oliveira:2012eh}.}	
\end{figure*}

{\it Lattice setup and algorithm.} In order to study the gluon propagator we use 50 configurations generated through importance sampling of the SU(3) Wilson action \cite{Cardoso:2011xu}. We opt for a symmetric lattice of size $L=32$ and $\beta = 6.0$, with associated lattice spacing $a = 0.1016(25)$ fm measured from the string tension~\cite{Bali:1992ru}. The simulated volume is therefore (3.25 fm)$^4$, large enough to resolve the onset of nonperturbative effects in the propagator's transverse form factor.  The values of $\xi$ chosen are $\xi$ = 0.1, 0.2, 0.3, 0.4, and 0.5. For comparison, we also report the Landau gauge, equivalent to $R_0$.
 
To gauge fix the configurations, we generate, for every lattice site $x$, $8$ real valued functions $\Lambda^m$ using the Gaussian probability distribution~\noeq{gaussian}; for the generation of the Gaussian distribution we used the standard Box-Muller algorithm. After combining these functions in the SU($3$) algebra element $\Lambda$, one proceeds to minimize the GF functional~\noeq{Rxifunct}. Notice that for each gauge configuration, we integrate over 50 different $\Lambda$'s. To reduce the correlations in the evaluation of the path integral, the $\Lambda$'s are generated independently for each gauge configuration. 

For minimization purposes we first tried to apply three different standard optimized techniques used in the Landau case~\cite{Cardoso:2012pv}: the Fast Fourier Transform -  accelerated steepest descent  (FFT), Over Relaxation (OVR) and Stochastic Relaxation (STR). Each one of these three techniques showed a typical GF success rate of around $\sim$ 75\% for $\xi=0.3$, with the convergence rate dropping to $\sim$ 40\% for $\xi=0.5$. Thus we opted to cycle through all convergence techniques when the procedure stalls. Indeed, by cycling through FFT, OVR and STR, for our hardest case of $\xi=0.5$, we increase the convergence success rate up to $\sim$90\%; for the remaining 10\% cases, restarting the combined algorithm, after performing finite random gauge transformations, leads to convergence for all cases
\cite{preparation}. \fig{fig:histo} compares the distribution obtained from the values of $\nabla\!\cdot\! A^m$ with the one expected for $\Lambda^m$ for a given configuration and a given color index ($m=4$); as one can appreciate the GF is within the precision defined above.

{\it Simulation results.} The lattice gluon two-point correlation function reads
\begin{align}
 \langle A^{m}_{\mu}(\widehat{q}\,) A^{n}_{\nu}(\widehat{q}\,') \rangle = \delta^{mn}\Delta_{\mu\nu}(q) L^4 \delta(\widehat{q}+\widehat{q}\,'),
 \label{lattprop}
\end{align}
where $\Delta_{\mu\nu}$ is given in~\1eq{Rxiprop}. The lattice momenta $\widehat q$ (used for Fourier transforms) and $q$ (the `continuum' momentum) are defined according to~\cite{Leinweber:1998im,Oliveira:2012eh},
\begin{align}
 q_{\mu}&=\frac{2}{a} \sin\frac{\widehat{q}_{\mu}}{2};& 
  \widehat{q}_{\mu}&=\frac{2\pi n_{\mu}}{ L},\quad n_\mu=1,2,\dots,L .
\end{align}

%---------------------------------------------------------------------------------------------------------------------
\begin{figure*}
%\includegraphics[width=0.95\columnwidth]{gl-trans}
%\hspace{40pt}
%\includegraphics[width=0.89\columnwidth]{ratios}
\includegraphics[scale=1]{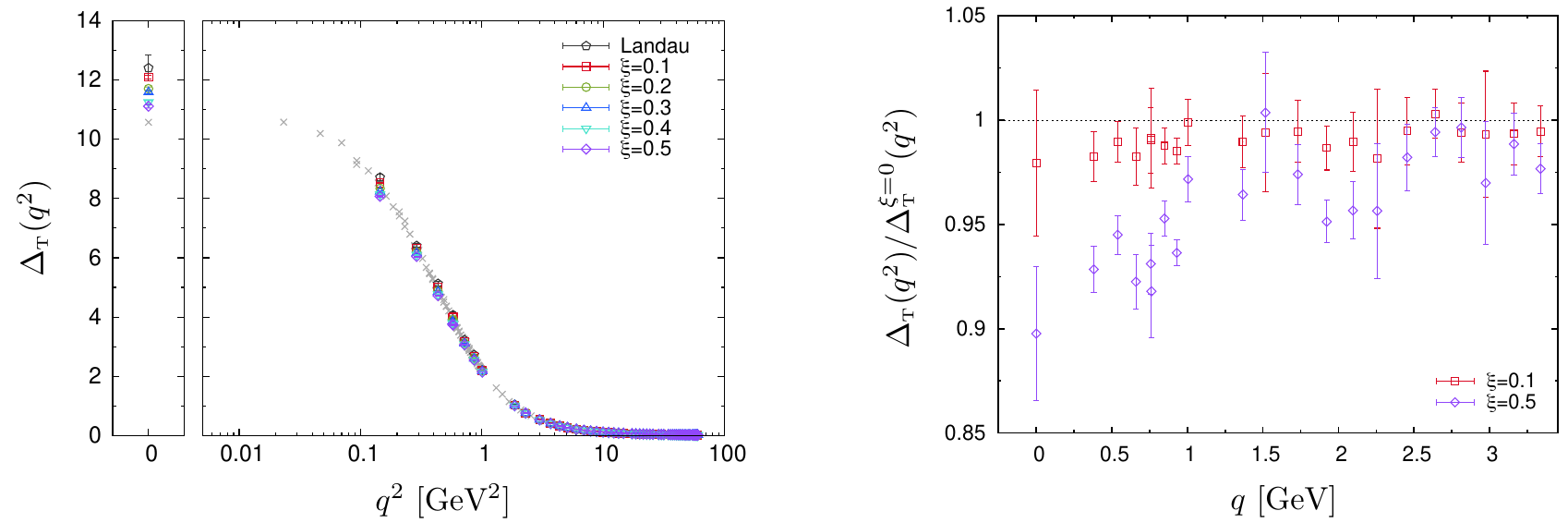}
\caption{\label{fig:gl-trans-ratios}
({\it Left}) 
The $R_\xi$ transverse propagator $\Delta_\s{\mathrm{T}}$ renormalized at $\mu=4.317$ [GeV]. The gray crosses (same as Fig. \ref{fig:gl-longit-trans} right~\cite{Oliveira:2012eh}) provide an estimate for the volume effects expected at $q^2=0$. ({\it Right}) The ratio $\Delta_\s{\mathrm{T}}(q^2)/\Delta^{\xi=0}_\s{\mathrm{T}}(q^2)$.
 }	
\end{figure*}

%---------------------------------------------------------------------------------------------------------------------
\begin{figure}
	\includegraphics[scale=0.65]{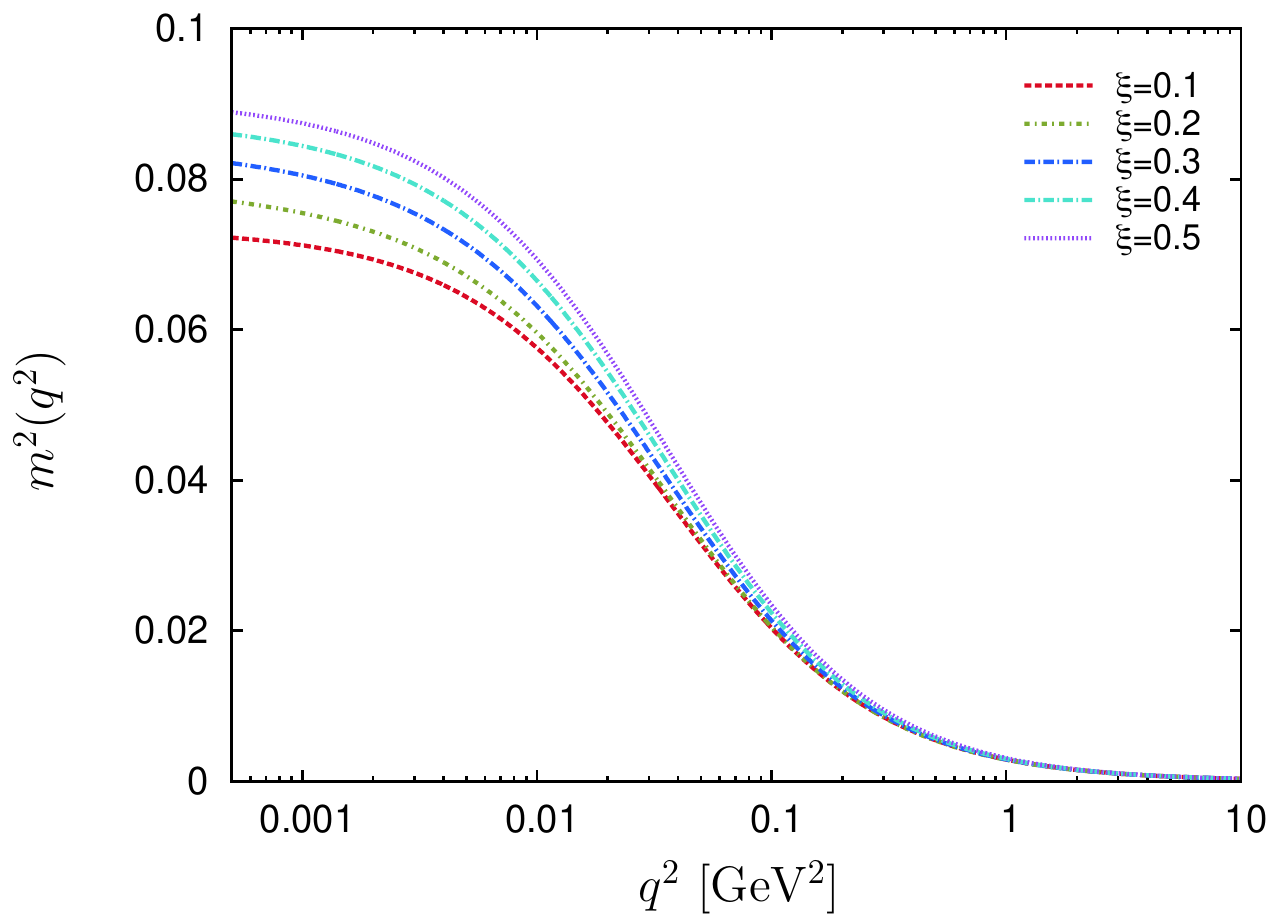}
	\caption{\label{fig:dress-mass} 
	The reconstructed dynamically generated gluon mass in the $R_\xi$ gauges.
	}
\end{figure}
  
From~\2eqs{Rxiprop}{lattprop} it follows that the transverse and longitudinal SU(3) propagator form factors can be estimated using,
\begin{align}
 \Delta_\s{\mathrm{T}}(q^2)&=\frac1{24L^4}\sum_{\mu,\nu,m}(\delta_{\mu\nu}-q_\mu q_\nu/q^2)\langle A^{m}_{\mu}(\widehat{q}\,) A^{m}_{\nu}(-\widehat{q}\,)\rangle,
 \nonumber \\
 \Delta_\s{\mathrm{T}}(0) &= \frac{1}{32L^4} \sum_{\mu,m}\langle A^m_\mu(0)  A^m_\mu(0)\rangle , \nonumber \\
 \Delta_\s{\mathrm{L}}(q^2)&=\frac1{8L^4}\sum_{\mu,\nu,m}q_\mu q_\nu/q^2\langle A^{m}_{\mu}(\widehat{q}\,) A^{n}_{\nu}(-\widehat{q}\,)\rangle .
\end{align}

To begin with, we show in ~\fig{fig:gl-longit-trans} the gluon dressing functions. Within statistical fluctuations, the longitudinal $q^2\Delta_\s{\mathrm{L}}(q^2)$ should be a constant function,  coinciding with the variance $\xi$ of the $\Lambda$ probability distribution. This is evidently true for all cases analyzed. In the right panel of~\fig{fig:gl-longit-trans} we plot the transverse $q^2\Delta_\s{\mathrm{T}}(q^2)$ for the different $\xi$ values studied, renormalized at $\mu=4.317$ GeV using a momentum subtraction scheme~\footnote{Within this scheme, for a given a set of data renormalized at the scale $\nu$, $\Delta(q^2;\nu^2)$, the data renormalized at a different scale $\mu$ can be obtained through the relation $\Delta(q^2;\mu^2)=\Delta(q^2;\nu^2)/\mu^2\Delta(\mu^2;\nu^2)$.}. Clearly, no significant deviation from the Landau gauge case is observed. 

Next, we turn our attention to the transverse form factor $\Delta_\s{\mathrm{T}}$  (\fig{fig:gl-trans-ratios}, left panel). As was already the case in the Landau gauge, one can see that the $R_\xi$ transverse propagators show an inflection point, implying that the associated spectral density is not positive definite; as this violates the quantum field theory axiom of reflection positivity, it is a manifestation of confinement~\cite{Cloet:2013jya}. In addition, they have a marked tendency to flatten towards the small momentum region, thus providing strong evidence that also in the $\xi\neq0$ case the behavior of the zero-momentum modes of the lattice gluon field are tamed by the dynamical generation of a (momentum-dependent) gluon mass. The data confirms an IR hierarchy such that $\Delta_\s{\mathrm{T}}$ (slightly) decreases for increasing values of the gauge fixing parameter~\cite{Cucchieri:2009kk,Giusti:2000yc}. This is better seen in~\fig{fig:gl-trans-ratios} (right) where we plot the ratio of the transverse propagator to the Landau gauge propagator $\Delta^{\xi=0}_\s{\mathrm{T}}$ as a function of the momentum for the two values $\xi=0.1$ and $\xi=0.5$; observing a maximum difference of about 10$\%$. 

{\it Comparison with continuum studies.} Fig. \ref{fig:gl-longit-trans} (right) turns out to differ from the effects observed in~\cite{Huber:2015ria} ({\it i.e.}, a dressing function in which the height of the peak rapidly increases and its location moves towards higher $q^2$ values with increasing $\xi$). Thus the effects reported are probably artefacts of the truncations employed.
On the other hand, our data agrees qualitatively with the Nielsen identities analysis of Ref. \cite{Aguilar:2015nqa}, where the results for the gluon propagator were interpreted in terms of the presence of a dynamically generated gluon mass.

We now briefly apply the analysis of Ref~\cite{Aguilar:2015nqa} to our $R_\xi$ lattice propagators.
First we write~\cite{Aguilar:2015nqa} $\Delta^{-1}_\s{\mathrm T}=q^2J_\s{\mathrm T}+m^2$, where, for small~$\xi$, one has,
\begin{align}
	m^2(q^2)&=\left[a(\xi)+c(\xi)\left(\frac{q^2}{\mu^2}\right)^{\!\!\xi}\log\frac{q^2}{\mu^2}\right]m^2_{\xi=0}(q^2) \, .
		\label{resmass}
\end{align}
To lowest order $a(\xi)=1+a_1\xi$, $c(\xi)=c_\s{\mathrm{NI}}\xi$, and $m^2_{\xi=0}$ is the Landau gauge dynamical gluon mass. 
By repeating the analysis of~\cite{Aguilar:2015nqa} using as input our Landau gauge $L=32$, $\beta=6.0$ data, we estimate~$c_\s{\mathrm{NI}}\approx 0.32$
~\footnote{Notice that the entire analysis of~\cite{Aguilar:2015nqa}  was performed using as input the (Landau gauge) lattice data of~\cite{Bogolubsky:2009dc}, and it would not be self-consistent to use the parameters (such as $c_\s{\mathrm{NI}}\approx0.13$) determined there.}. Next, we simply fit $a_1$ by requiring that the resumed mass~\noeq{resmass} equals the value of $\Delta_\s{\mathrm T}^{-1}(0)$ for the corresponding value of the GF parameter $\xi$ and solving the renormalization group improved gluon mass equation~\cite{Aguilar:2014tka} (yielding  $m^2_{\xi=0}$). We obtain $a_1 \approx 0.26$, and the resulting mass is plotted in~\fig{fig:dress-mass}. 

{\it Conclusions and outlook.} In this paper the lattice SU(3) gluon propagators in $R_\xi$ gauges is computed for large $\xi$'s and for a lattice volume large enough to access the IR dynamics. From the numerical point of view, the most intensive task was the gauge fixing due to the large number of GF's required and algorithmic issues. At least for the set of parameters simulated here, a proper combination of various methods solved the minimisation problem associated with the GF in $R_\xi$ gauges; this will be presented in detail elsewhere \cite{preparation}. 

Our $R_\xi$ propagators show very similar characteristics to the one in the Landau gauge: an inflection point in the few hundreds MeV region followed by saturation to a finite value in the IR. Comparing with very recent continuum analytic studies, our propagators are in agreement with Ref.~\cite{Aguilar:2015nqa}, which allowed us to estimate the dynamically generated $R_\xi$ mass. This analysis suggests that dynamical gluon mass generation is a common feature of all $R_\xi$ gauges in SU(3) Yang-Mills theories.

{\it Acknowledgments.}
The authors acknowledge the use of the computer cluster Navigator, managed by the Laboratory for Advanced Computing, at the University of Coimbra, of the GPU servers of PtQCD, supported by CFTP, FCT and NVIDIA; AC33 and Hybrid GPU servers from NCSA's Innovative Systems Laboratory (ISL). Simulations on Navigator were performed using Chroma \cite{Edwards:2004sx} and PFFT  \cite{Pippig:2011} libraries. We thank A.~C.~Aguilar for providing the data needed for the continuum analysis, and J.~Papavassiliou for a critical reading of the manuscript.  P.~B. thanks  the ECT* Centre for hospitality and CFTP, grant FCT UID/FIS/00777/2013, for support. The work of N.~C. is supported by NSF award PHY-1212270. P.~J.~S. acknowledges the support of FCT through the grant SFRH/BPD/40998/2007. 

%\bibliography{bibliography}

%merlin.mbs apsrev4-1.bst 2010-07-25 4.21a (PWD, AO, DPC) hacked
%Control: key (0)
%Control: author (8) initials jnrlst
%Control: editor formatted (1) identically to author
%Control: production of article title (-1) disabled
%Control: page (0) single
%Control: year (1) truncated
%Control: production of eprint (0) enabled
%

\end{document}